\documentclass{article}

\usepackage[utf8]{inputenc}
\usepackage{graphicx}
\usepackage{amsfonts}
\usepackage{subcaption}
\usepackage{url}
\usepackage{float}
\usepackage[normalem]{ulem}
\usepackage[labelfont=bf, font=scriptsize]{caption}
\usepackage{placeins}
\usepackage{authblk}
\usepackage{comment}
\usepackage[a4paper, total={6.3in, 10in}]{geometry}
\usepackage[colorlinks=true, citecolor=blue, filecolor=blue, linkcolor=black]{hyperref}
\usepackage{bbm}
\usepackage{color}
\usepackage{epstopdf}
\usepackage{comment}
\usepackage{multicol}
\usepackage{amsmath} 
\usepackage{color}
\usepackage[utf8]{inputenc} % allow utf-8 input
\usepackage[T1]{fontenc}    % use 8-bit T1 fonts
\usepackage{url}            % simple URL typesetting
\usepackage{booktabs}       % professional-quality tables
\usepackage{amsfonts}       % blackboard math symbols
\usepackage{nicefrac}       % compact symbols for 1/2, etc.
\usepackage{microtype}      % microtypography
\usepackage{lipsum}
\usepackage{placeins}
\usepackage{xcolor,colortbl}
\usepackage{amsmath}
\usepackage[numbers,sort&compress]{natbib}
\usepackage{lastpage,fancyhdr,graphicx}
\title{\textbf{Quantifying agent impacts on contact sequences in social interactions}}
\author[1,2,*]{Mark M. Dekker}
\author[3]{Tessa F. Blanken}
\author[3]{Fabian Dablander}
\author[1,4]{Jiamin Ou}
\author[3]{Denny Borsboom}
\author[1,2]{Debabrata Panja}
\affil[1]{\small{Department of Information and Computing Sciences, Utrecht University, Princetonplein 5, 3584 CC Utrecht, The Netherlands}}
\affil[2]{Centre for Complex Systems Studies, Utrecht University, Minnaertgebouw, Leuvenlaan 4, 3584 CE Utrecht, The Netherlands}
\affil[3]{Department of Psychological Methods, University of Amsterdam, Nieuwe Achtergracht 129-B, 1018 VZ Amsterdam, The Netherlands}
\affil[4]{Department of Sociology, Utrecht University, Padualaan 14, 3584 CH Utrecht, the Netherlands}

\affil[*]{m.m.dekker@uu.nl}

\begin{document}
\maketitle

%% ============================================================================ %%
%% Abstract
%% ============================================================================ %%

\begin{abstract}
\noindent Human social behavior plays a crucial role in how pathogens like SARS-CoV-2 or fake news spread in a population. Social interactions determine the contact network among individuals, while spreading, requiring individual-to-individual transmission, takes place on top of the network. Studying the topological aspects of a contact network, therefore, not only has the potential of leading to valuable insights into how the behavior of individuals impacts spreading phenomena, but it may also open up possibilities for devising effective behavioral interventions. Because of the \textit{temporal} nature of interactions --- since the topology of the network, containing who is in contact with whom, when, for how long, and in which precise sequence, varies (rapidly) in time --- analyzing them requires developing network methods and metrics that respect temporal variability, in contrast to those developed for static (i.e., time-invariant) networks. Here, by means of event mapping, we propose a method to quantify how quickly agents mingle by transforming temporal network data of agent contacts. We define a novel measure called \textit{contact sequence centrality}, which quantifies the impact of an individual on the contact sequences, reflecting the individual's behavioral potential for spreading. Comparing \textit{contact sequence centrality} across agents allows for ranking the impact of agents and identifying potential `behavioral super-spreaders'. The method is applied to social interaction data collected at an art fair in Amsterdam. We relate the measure to the existing network metrics, both temporal and static, and find that (mostly at longer time scales) traditional metrics lose their resemblance to \textit{contact sequence centrality}. Our work highlights the importance of accounting for the sequential nature of contacts when analyzing social interactions.
\end{abstract}

%% ============================================================================ %%
%% Main text - Section 1
%% ============================================================================ %%
%% ============================================================================ %%
\section{Introduction}
\label{sec:1}
%% ============================================================================ %%

Human behavior plays a central role in generating patterns of interaction that allow for the spreading of a great variety of entities --- from rumors \cite{Choi2020} to viruses \cite{Stehle2011b} to memes \cite{Kotsakos2015}. Most recently, the importance of such patterns of interaction has been borne out by the COVID-19 crisis. A virus like SARS-CoV-2, which caused the COVID-19 pandemic, spreads through physical contact, or through aerosols that have a finite range; for the virus to be transmitted, people will have to have make close contacts with each other. It is for this reason that behavioral interventions, aiming to break the chains of close human contacts, were our only weapon against COVID-19 for the better part of 2020 and remain so until vaccine coverage is sufficient.

As such, the pandemic underscored the need for better models and measures of human behavior and the contact patterns it generates. On the technological side, recording individual-level contacts through wearable sensor technology has recently seen considerable development that render the study of contact patterns scientifically feasible  \cite{Genois2018, Cattuto2010, Stehle2011}. For example, albeit still imperfect, sensor accuracy is beginning to approach the level and scale needed to inform epidemiology and to model interventions \cite{Masuda2013, Zhang2021, Li2018}. In contrast, on the methodological side, we still lack adequate tools to assess, represent, and model the myriad ways in which human behavior generates patterns of individual-level contacts. This limits progress in many disciplines that require such information, but especially hampers progress along the behavior-epidemiology interface that is so important in improving our preparedness to deal with the current and future pandemics. This paper develops methodologies that can begin to fill this gap, by exploiting connections among improved empirical assessments of contact patterns and novel quantitative metrics that are sufficiently advanced to analyze behavioral contact patterns. In particular, it focuses on quantifying the role of each individual in the overall contact patterns, reflecting the agent's impact on the potential of (e.g., epidemiological) spreading in a system.

To achieve these goals, we utilize a network approach in which patterns of contacts which are generated by individual behavior are mapped on to a contact network. In such a network, each individual corresponds to a node and a link between two nodes represents a contact between two individuals. The advantage of using a network approach is that a general spreading process of some entity (of which pathogen spreading in epidemiology is one, where the concerned pathogen is the entity) can be analyzed by means of a two-level description --- the ``infrastructure'', and the process on top of it: (a) a contact network, borne out of people's behavior in social interactions that constitutes the infrastructure, and (b) a pathogen spreading process occurring within a population, which then becomes a (dynamical) process that uses this infrastructure as pathways to transmit from an infected to a susceptible agent. From a network science perspective, it then becomes natural to expect that the topological properties of the interaction networks --- the infrastructure --- can be used to analyze the {\it potential\/} of spreading dynamics. In addition, this approach allows us to analyze the relative importance of different individuals' potential of spreading by analyzing the effect that removing the individual would have on the network structure. Individuals who contribute greatly to the topological aspects of the network may then represent potential targets of intervention. (In contrast, social distancing is a non-pharmaceutical intervention measure against SARS-CoV-2 transmission by means of globally disrupting the infrastructure).

It is important to note, however, that spreading dynamics can take on many forms: infectious diseases will spread differently from rumors or political opinions, which are yet different from spreading of delays in network transport \cite{Dekker2021plos}. Nevertheless, the common element they all share is (deterministic or stochastic) transmission upon contact, and spreading is therefore dependent on the topology of the contact network: who interacts with whom, for how long and in which precise sequence. For example, if a person $A$ spreads the rumor, a person $B$ can only spread the rumor to person $C$ if (s)he has been in contact with $A$ before the contact with $C$. Tracking these sequences of interactions, therefore, provides the basis of any spreading phenomenon. While the work presented here was developed in the context of modeling behavior during the COVID-19 pandemic, the importance of better models for human behavior in generating contact networks should therefore be seen as generic.

Before a satisfactory analysis of this process can be found, however, several problems that arise in the currently used network analytic techniques (for evaluating the structure of contact networks and the role of different individuals in generating them) must be addressed. First and foremost, in reality, a contact network is dynamic, i.e., the topology of a contact network is changes over time. In Network Science such networks are called {\it temporal networks\/}. Many existing methods that analyze contact networks, empirically obtained through wearable sensors \cite{Blanken2020}, treat contact networks as static, since, for instance, recorded contacts may not be adequately logged or the analysis needs to be simplified by collapsing networks over the time domain. From a Network Science perspective, this is in fact reminiscent of one way to reduce the complexity of a temporal network by downscaling it to a static one: a network with a time-invariant topology, obtained through aggregation of all edges over a certain time interval \cite{Li2020, Isella2011, Riolo2001, Masuda2013, Stehle2011}. However, information on who interacts with whom, for how long and in which precise sequence gets lost in such aggregations. Static representations therefore do not reflect the dynamic aspects of human behavior: changing interactions, structures and sequences \cite{Li2017, Li2020, Mucha2010, Genois2018, Centola2010, Schlapfer2014, Kiti2019, Cattuto2010} have been shown to have large impacts on various dynamical processes \cite{Mucha2010, Perra2012, Scholtes2014, Peixoto2017}. Static representations miss a crucial opportunity to intervene at the contact network level, e.g., by targeting interventions at specific points in the time domain. Secondly, while there are \textit{static} topological metrics available to quantify the impact of agents on spreading mechanisms, static network techniques cannot differentiate between agents whose contacts are structured in the same way but located at different time points. Examples of such topological metrics are degree, closeness or betweenness centrality \cite{Dezso2002, Pastor2002}, $k$-shell decomposition algorithm \cite{Kitsak2010}, methods based on clustering and the topology potentials \cite{wang2015}, improved coreness centrality and eigenvector centrality \cite{Ahajjam2018}, influential individuals \cite{Zhang2019}.

More prudent (existing) approaches to deal with the complexity of temporal contact networks entail setting a threshold to filter out the non-essential edges or edges that exist only by chance: for example, Grabowicz {\it et al.\/} used a simple threshold based on the number of events (i.e., count of edges across time) between two nodes \cite{Grabowicz2014}; Kobayashi defined a temporal null model to identify pairs of nodes having more interactions than expected given their activities \cite{Kobayashi2019}; and in yet another thread of study, Mellor {\it et al\/} introduced the temporal event graph (TEG), which uses events (interactions between two individuals) as nodes and shared event attendees as edges in a directed graph \cite{Mellor2017, Mellor2019}. Some of these efforts focus on only part of the complexity of temporal networks and provide possible ways to identify structures \cite{Grabowicz2014, Kobayashi2019}, communities \cite{Genois2018, Peixoto2017, Peel2017} and quantify connectivity \cite{Mellor2019, Badie2020, Torricelli2020}. Other related papers on the topic discuss temporal motifs \cite{Saramaki2019} and percolation in light of weighted temporal event graphs \cite{Kivela2018}, or propose an event embedding technique to obtain a low-dimensional representation of temporal networks \cite{Torricelli2020}. However, in the context of spreading phenomena, these metrics are not meant for quantifying agent impacts on contact chains in human interactions (and thereby the spread of a concerned entity). We also use event mapping, which we explain in Sec.~\ref{sec2-1}.

A more principled approach to unraveling behavioral heterogeneity and the impact of individual actors, which we develop in this paper, is to consider the full dynamics of people's contacts in the form of a temporal network. Specifically, we provide a way forward by developing (a) a systematic methodology for representing people's (dynamical) contact networks as event graphs, and (b) techniques to analyze the role that individuals play in generating these network structures. The latter information can be used to assess individuals' impacts on the contact sequence \textit{while respecting (temporal) sequence of contacts}. 

This paper is structured as follows. We start by introducing the methodology to extract an individual's impact on contact sequences in a system in Sec.~2, for which we need to introduce a reconstruction of the temporal graph using `event mapping'. In Sec.~3, we apply the methodology to data collected during an art fair in Amsterdam in August 2020, where we also provide experimental details and discuss the evolution of the most important metrics over time, as well as relate them to the existing network metrics like contact degree. In Sec.~4, we discuss these findings and relate them to the broader discussion on spreading phenomena in networked systems. We end with concluding remarks in Sec.~5.

%% ============================================================================ %%
%% Main text - Section 2
%% ============================================================================ %%
%% ============================================================================ %%
\section{Methods}
\label{sec:2}
%% ============================================================================ %%

In this section, we introduce the mapping of temporal network data into an event graph in order to facilitate the identification of agent impacts on a sequence of contacts.

%% ==================================== %%
\subsection{Event mapping of contact sequences\label{sec2-1}}
%% ==================================== %%

\begin{figure}[!h]
    \centering
    \includegraphics[width=1\linewidth]{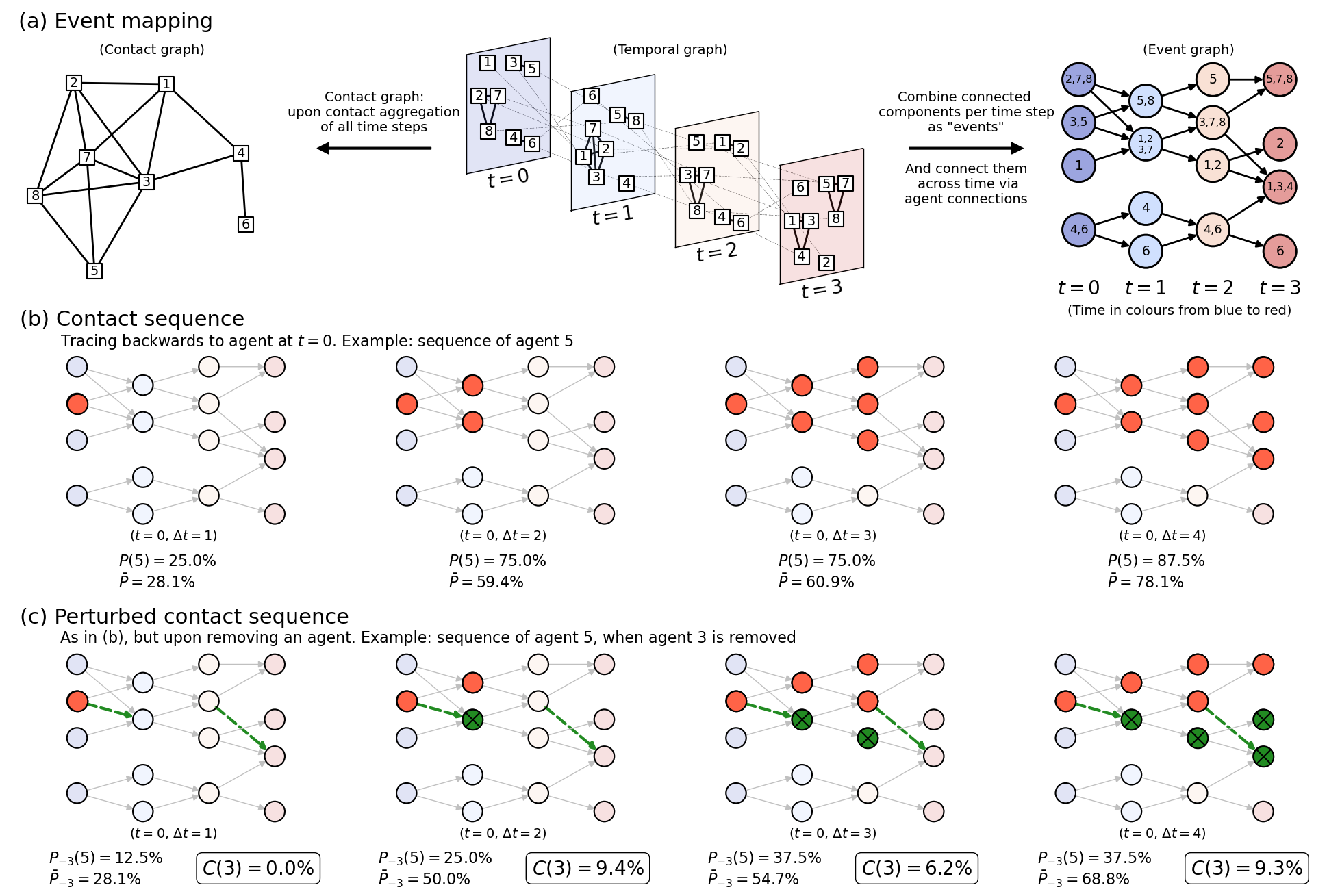}
    \caption{Illustration of the event mapping and an individual's contact sequence centrality using a toy example. All throughout this figure, individual agents are in squares (and numbers), events are in circles. \textbf{Panel (a)}: In a temporal network, the time evolution of the contact graph (middle) captured in a sequence of snapshots. Contact data over the four snapshots can be aggregated to obtain a static (and aggregated) contact network (left), but in doing so, the sequential information is lost. We term the connected network components within each snapshot `events', they are seen as nodes in an event graph (right). In an event graph the event-to-event trajectories of agents are the links. Colors of circles and snapshots indicate time from blue to red. \textbf{Panel (b)}: Example of a contact sequence when tracing all events and subsequent agents back to agent 5 at $t=0$. Starting time $t$, time elapsed $\Delta t$, the percentage unique agents [$P(5)$], traceable to 5 and the average percentage of agents in the contact sequence $\bar{P}$ are indicated.
    \textbf{Panel (c)}: Same as (b), but after removal of agent 3, resulting in a `perturbed' contact sequence of agent 5. The (now absent) links that were associated with agent 3 are in dashed green arrows, and the events that are consequentially not traceable to 5 anymore are colored green. The associated (lowered value of) $P_{-3}(5)$ is shown. Averaging over all agents provides $\bar{P}_{-3}$, and allows us to calculate the \textit{Contact Sequence Centrality} of agent 3: $C(3) = \bar{P} - \bar{P}_{-3}$, which is clearly marked in the bottom panel. For abbreviation, we excluded the $t$ and $\Delta t$ between brackets when showing values of $P$, $\bar{P}$ and $C$.}
    \label{fig:1}
\end{figure}

We track the contact sequences of the agents in the following manner. In Fig.~\ref{fig:1}(a) we show an example of a (randomly generated) temporal system with eight agents and four time steps $t\in\{0, 1, 2, 3\}$ in three types of graphs. In the center, the temporal graph is shown where agents are denoted in squares. The contacts among the agents at every snapshot are denoted by black links, while the agents' movements across snapshots are shown by gray links. To the left of the panel, we show the corresponding aggregated \textit{contact network} over the entire time range, wherein two agents are connected by an edge if they have been in contact at any of the four time steps: the aggregated contact network therefore does not distinguish any sequence between the interactions.

To the right of Fig.~\ref{fig:1}(a), we combine connected components within time snapshots into single circular nodes, which we refer to as \textit{events} \cite{Dekker2021temp}, resulting in yet another graph: the \textit{event graph}. We use a \textit{non-weighted} version of event mapping, where links between events are not weighted by the number of crossing participants. Events may consist of either single or more agents, as marked by the numbers indicated inside each node. The events are linked by the `re-usage' of agents; this is exactly how every agent can act like a spreader between consecutive events, which makes event mapping a natural choice for describing the dynamical evolution of a spreading phenomenon. For example, at $t=0$, agent 5 is interacting with agent 3. From there, they go their separate ways: at $t=1$ agent 5 interacts with agent 8, and agent 3 interacts (in a larger group) with agents 1, 2 and 7. Indeed, the event (3, 5) at $t=0$ links to both (5, 8) and (1, 2, 3, 7) at $t=1$. Using these events, one can convert the (temporal) network snapshots to an event graph (directed in time), shown at the right of Fig.~\ref{fig:1}(a). Not only does this \textit{event mapping} collapse complex temporal network data into a single directed graph, but most importantly, it also preserves the sequences of contacts. [An important question is to what extent the information for spreading phenomena is already embedded in the static contact graph, and when it is important to do event mapping. This is a question we address multiple times in this paper, by comparing the results of our metrics (defined below) to properties of the static contact network.]

The event graph allows for an intuitive analysis of which (agent contacts in) events in later time steps are eventually traceable back to the an event in earlier time steps via both direct and indirect connections. An example of such a contact sequence is given in Fig.~\ref{fig:1}(b). There, we track the events that are linked to agent 5 at $t=0$ (colored red), over four sequential time snapshots. At $t=0$, only the event where agent 5 itself is involved is colored red [i.e., agents (3,5)]. Over time, `secondary' events become colored red in the event graph, meaning that in due course of time, progressively more events, and the agents contained therein, can be traced backwards to agent 5 at $t=0§$. In the end, for most applications, how many agents can be traced backwards to agent 5 at $t=0$ is the most relevant quantity (rather than the number of events, i.e., the number of red dots). This prompts us to define $P(j, t, \Delta t)$:

\begin{eqnarray}
P(j, t, \Delta t) &=& \frac{\text{Number of agents traceable to agent $j$ at time $t$ within time interval $[t, t+\Delta t$]}}{N},
\end{eqnarray}

where $N$ is the total amount of agents. In other words, $P(j, t, \Delta t)$ is the fraction of {\it unique\/} agents at time $t+\Delta t$ that can ultimately be linked backwards to agent $j$ at time $t$ --- clearly, $P(j, t, \Delta t)$ is a function of the starting time $t$ and elapsed time $\Delta t$. In Fig.~\ref{fig:1}(b), we show $P(5, 0, \Delta t)$ for different values of $\Delta t$, using agent 5 as `agent zero': at $\Delta t = 0$, $P(5, 0, 1) = \nicefrac{2}{8} = 25\%$ since at $\Delta t=0$ two (3 and 5 itself) out of eight agents are linked to 5 at $t=0$. Similarly, $P(5, 0, 2) = \nicefrac{6}{8} = 75\%$ since there are six agents (1, 2, 3, 5, 7 and 8) that can be traced backwards to agent 5 at $t=0$, and so on. Clearly, a high value of $P(j,t,\Delta t)$ for low $\Delta t$ would indicate fast mingling of the agents in the time interval $(t,t+\Delta t)$, making the system highly prone to spreading phenomena from the perspective of starting with agent $j$ at $t=0$. Conversely, a low value of $P(j,t,\Delta t)$ for high $\Delta t$ would imply the opposite. It is obvious that for fixed $t$, $P(j,t,\Delta t)$ is a monotonically increasing function of $\Delta t$, while for fixed $\Delta t$ and a different initialization time $t$, $P(j, t, \Delta t)$ may vary and therefore possibly result in lower values with increasing $t$.

In panel (b), we use agent 5 as a mere example to illustrate the calculation of $P(j=5, t, \Delta t)$. A more representative quantity is obtained by taking the average over all agents $j$:

\begin{eqnarray}
\bar{P}(t, \Delta t) &=& \frac{1}{N}\sum_j^NP(j, t, \Delta t)
\end{eqnarray}

the value of which is also indicated in Fig.~\ref{fig:1}b. The value of $\bar{P}$ says something about the (unperturbed) contact sequences: high values indicate that spreading to many agents within ($t, t+\Delta t$) is likely, and vice versa. We aim to identify the impact of each agent on the overall contact sequences. We measure this by looking at the $\bar{P}$ value when we remove an agent -- the intuition being that the resulting change in $\bar{P}$ reflects the impact of the removed agent. Again, we start with the example of the contact sequence of agent 5 (i.e, $P(5)$ instead of $\bar{P}$). Now, to assess the impact of \textit{another} agent -- agent 3 -- on agent 5's contact sequence, we \textit{remove} this agent from the system and show what it does to agent 5's contact sequence in Fig.~\ref{fig:1}(c). The removal results in the absence of important links (denoted in green dashed lines) and, consequently, in limiting the contact sequence's reach (i.e., excluding the green nodes). The proportion of agents in agent 5's contact sequence is now written as $P_{-3}(5, t, \Delta t)$, where the subscript $-i$ refers to the removal of agent $i$. In particular, we have $P_{-3}(5, 0, 2) = \nicefrac{2}{8} = 25\%$, which is lower than $P(5, 0, 2)$, since the large event comprised of agents 1, 2, 7 and the removed 3 cannot be traced back to agent 5 at $t=0$ anymore. More generally, for any agents $i, j$ the relation $P_{-i}(j,t,\Delta t) \leq P(j, t, \Delta t)$ holds, because the contact sequence is either equal or shrunk by the removal of agent $i$.

The impact of an agent on the contact sequence is naturally quantified by the difference in $\bar{P}$ between the unperturbed [panel (b)] and the perturbed event graph. For the impact of agent 3, the latter is shown in panel (c). Averaging over the contact sequences of each agent $j$, we obtain an expression for the impact of an individual agent on the contact sequence which we refer to as \textit{contact sequence centrality}:

\begin{align}
   C(i, t, \Delta t) &= \frac1N\left[\sum_j P(j, t, \Delta t) - \sum_j P_{-i}(j, t, \Delta t)\right]\nonumber \\
     &= \bar{P}(t, \Delta t) - \bar{P}_{-i}(t, \Delta t).
     \label{e2}
\end{align}

A high $C(i, t, \Delta t)$ indicates that the removal of agent $i$ decreases the fraction of agents in an average contact sequence sharply, which reflects an important role of $i$ in connecting contact sequences. This way, $C(i, t, \Delta t)$ becomes an attribute of agent $i$. It might just reflect that agent $i$ itself has had a lot of contacts in the interval $(t,t+\Delta t)$ (i.e., the degree of agent $i$ in the aggregated contact graph is high), but that is not necessarily the case. It may also be that agent $i$ serves as the intermediary between two larger `bubbles' or temporal communities, which will become disjointed if agent $i$ is removed from the system. Example values of $C(3, t, \Delta t)$ are shown in Fig.~\ref{fig:1}(c) for the case of agent 3.

%% ==================================== %%
\subsection{Other agent-based topological network metrics}
%% ==================================== %%

It is important to assess whether the contact sequence centrality $C$ merely reflects static properties of the network topology that could have been found more easily using traditional methods, or whether it provides us with new information regarding an agent's role in the system. This prompts us to compare $C$ values to various other agent-based metrics that have been around in the literature. Some of them are derived from the contact network like the one shown Fig.~\ref{fig:1}(a) on the left, obtained by aggregating all the temporal network snapshots into one single, unweighted, undirected contact network within a certain (or the whole) time interval. The \textit{degree} of an agent $i$ equals the amount of links this agents has, i.e. the amount of unique agents $i$ had been interacting with. Similarly, the \textit{contact betweenness} is the betweenness centrality in the aggregated contact network; betweenness centrality is defined as the fraction of all shortest paths (i.e., between all unique pairs of agents in the aggregated network) that passes through an agent. As for event maps of temporal networks, the \textit{number of events} at any time snapshot is the amount of events of sizes larger than 1 (i.e., `non-individual events') the agent attends. The \textit{average event size} determines the average amount of attendants of events in general.

Finally, for temporal networks, many non-static quantities have been developed, as variants of their static counterparts, in order to study the temporal structural properties of the system  \cite{hafiene2020influential, masuda2020guide}. While they have been developed for degree and closeness as well \cite{Kim2012}, we choose to compare $\Delta \bar{P}$ to two types of betweenness other than the aforementioned contact betweenness. To explain the difference, we refer to the graphs shown in Fig.~\ref{fig:1}(a). On the left, we see the contact graph, in which the contact betweenness (already mentioned above) is defined. On the right, we see an event graph, where the nodes are events. The betweenness -- i.e. fraction of shortest paths going through them -- of nodes in this graph is computed, and when taking the average of all events where any agent $i$ is participating, we obtain the value of what we refer to as the \textit{event betweenness} of this agent. Likewise, we can focus on the network in the middle: the temporal graph. Nodes here are agents at a specific time snapshot. The betweenness in this graph is computed, and then we again average over all nodes in this graph belonging to each agent individually, which results in what we refer to as the \textit{temporal betweenness} of an agent. The event betweenness and temporal betweenness are related, but not necessarily the same, as finer network structures within events are not incorporated in the event betweenness (while they are in temporal betweenness), and the event betweenness calculates shortest paths between pairs of events rather than pairs of agents.

Because of computational limitations, we choose to calculate temporal betweenness and event betweenness in time intervals smaller than the full data sets (i.e., we take ten subsequent 4-min time intervals). This results in multiple values per agent (each belonging to one of these intervals), of which the average is reported in later usage of these metrics.

%% ============================================================================ %%
%% Main text - Section 3
%% ============================================================================ %%
%% ============================================================================ %%
\section{Application to human interaction data}
\label{sec:3}
%% ============================================================================ %%

In this section, we perform event mapping and contact sequence calculations on human interaction data measured at an art fair in Amsterdam. After describing the experiment details in Sec. 3.1, we discuss the resulting $\bar{P}$ and $C$ values in Sec. 3.2, and compare the latter to other metrics in Sec. 3.3.

%% ==================================== %%
\subsection{Experiment and sensor data}
%% ==================================== %%

\begin{figure}[!h]
    \centering
    \includegraphics[width=0.8\linewidth]{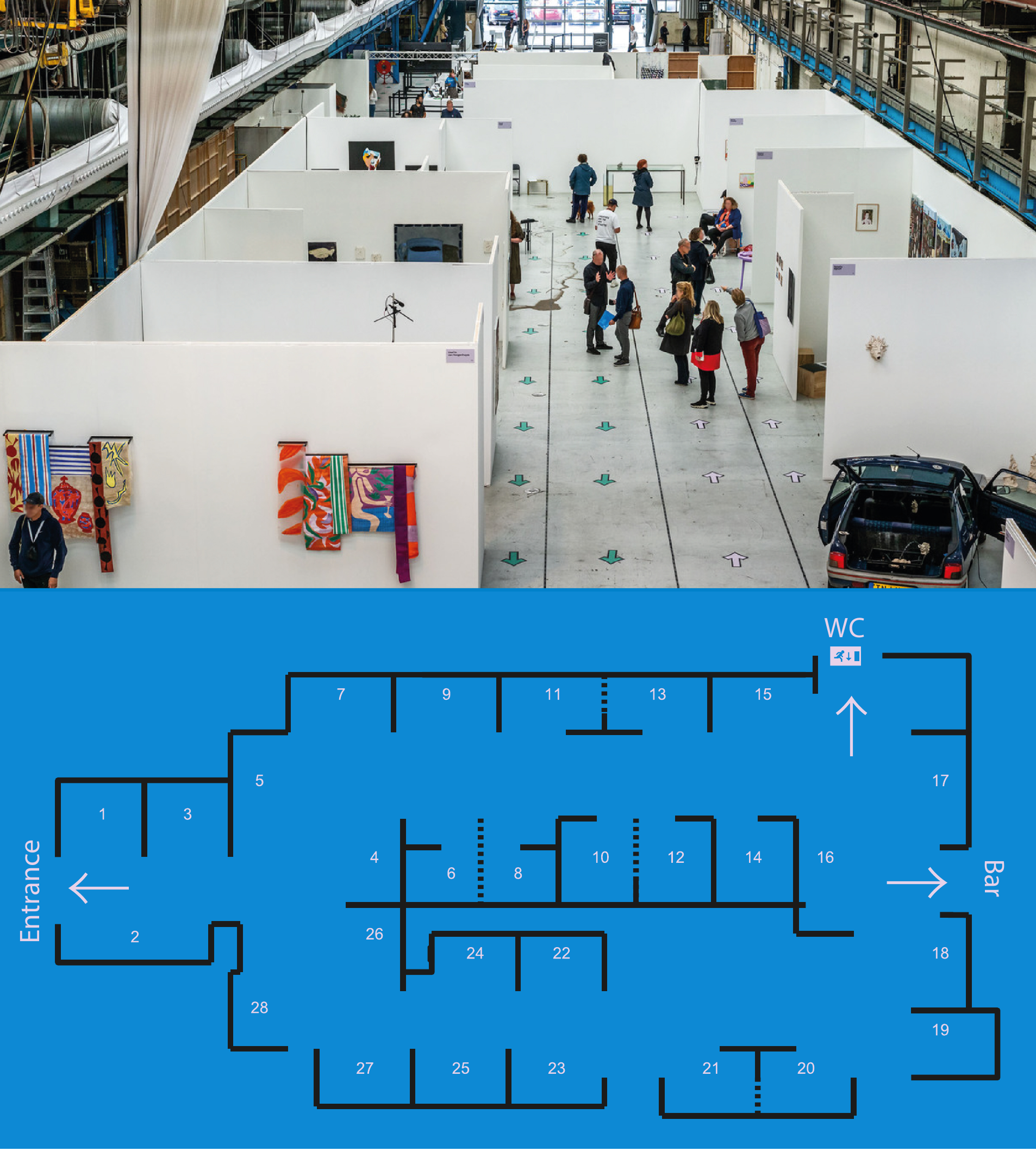}
    \caption{The art fair on Amsterdam on the first day with bidirectional walking directions (top) and schematic layout of the location (bottom). The art fair consisted of 28 stands, spanning 1,080m$^2$. Visitors entered on the left of the plan in the bottom panel, where also the access point was placed. Figure adapted and reprinted with permission from Ref. \cite{Tanis2020}.}
    \label{fig:2}
\end{figure}

In August 2020, after the first peak of COVID-19 infections was decaying in the Netherlands, right before the start of the second wave of infections, an art fair was organized in Amsterdam with the goal of assessing the effectiveness of physical distancing interventions. For an illustration of the art fair, see Fig.~\ref{fig:2}. During this three-day art fair, different interventions to promote physical distancing were implemented: walking directions, face masks, and a buzzer-notification (a buzzing sound when coming within 1.5m of another visitor). The three day art fair was split into eleven time slots and the aforementioned interventions varied across these time slots. Each visitor was asked to wear a sensor that recorded the distance to the sensors of other visitors within line of sight (max 30m) using ultra wide band technology. Whenever two sensors were within 1.5m of one another the opposing tag ID was registered and the contact was logged locally (at a 1-second resolution), which was sent to a database via access points, placed near the entrance of the art fair. Researchers from the University of Amsterdam collected the questionnaire and sensor data and have published this in previous work \cite{Tanis2020, Blanken2020}. The ethics review board of the University of Amsterdam (2020-CP-12488) approved data collection, and all participants provided informed consent before participating. All personally identifiable information used to link the questionnaire and sensor data has been destroyed. All methods were performed in accordance with the relevant guidelines and regulations.

From time to time, in most experiments, the interaction registration was hampered by not synchronizing to the access points well, altering some of the time stamps. Because the time stamps are crucial to the analysis in this paper, we chose to limit the analysis to parts of the experiments that did not contain such gaps (see Appendix A for a supplementary figure on choosing the correct parts of the data).

Table~\ref{tab:exps} shows an overview of the experiments, their (filtered) duration, conditions and the resulting amount of agents. More details can be found in Ref. \cite{Tanis2020}.

\begin{table}[!h]
\resizebox{\textwidth}{!}{%
    \centering
    \begin{tabular}{l|ccccccc}
        \hline
        \textbf{Exp} & \textbf{Duration} & \textbf{Walking} & \textbf{Supplementary} & \textbf{SDS setting} & \textbf{\# Agents} & \textbf{Avg. degree} & \textbf{Avg. degree}\\
        & $[\text{min}]$ & \textbf{direction} & \textbf{intervention} & & & & \textbf{per agent}\\
        \hline
        Exp 2 & 19 min & Bidirectional & Facemask & No feedback &  98 & 10.2 & 0.10\\
        Exp 3 & 25 min & Bidirectional & None & No feedback & 95 & 5.1 & 0.05\\
        Exp 4 & 50 min & Bidirectional & Buzzer & Buzzer after 2 sec & 88 & 18.1 & 0.21\\
        Exp 6 & 55 min & Unidirectional & None & No feedback & 130 & 19.6 & 0.15\\
        Exp 7 & 38 min & Unidirectional & Buzzer & Buzzer immediately & 88 & 7.9 & 0.09\\
         &  &  &  & stops after 2 sec & & & \\
        Exp 9 & 85 min & No direction & Buzzer & Buzzer immediately & 109 & 17.3 & 0.16\\
         &  &  &  & stops after 2 sec &  & & \\
        Exp 10 & 42 min & No direction & Buzzer & Buzzer immediately & 120 & 11.7 & 0.10\\
         &  &  &  & persists after 2 sec & & & \\
        Exp 11 & 44 min & No direction & None & No feedback & 89 & 17.4 & 0.20\\
        \hline
    \end{tabular}}
    \caption{Details of the conditions and attributes of the experiments. In the second column (`Duration') we note the filtered window durations, rather than the full experiment times. Event amounts and link frequencies are determined using only events with sizes larger than 1. Number of agents may differ from that in Ref. \cite{Tanis2020} because we focus on only a part of each experiment (see Fig.~A.1). The average degree and the average degree per agent have been calculated from the aggregated contact network, aggregated over the full valid duration of the experiments.}
    \label{tab:exps}
\end{table}

%% ==================================== %%
\subsection{Evolution of $\bar{P}$ and $C$}
%% ==================================== %%

\begin{figure}[!h]
    \centering
    \includegraphics[width=1\linewidth]{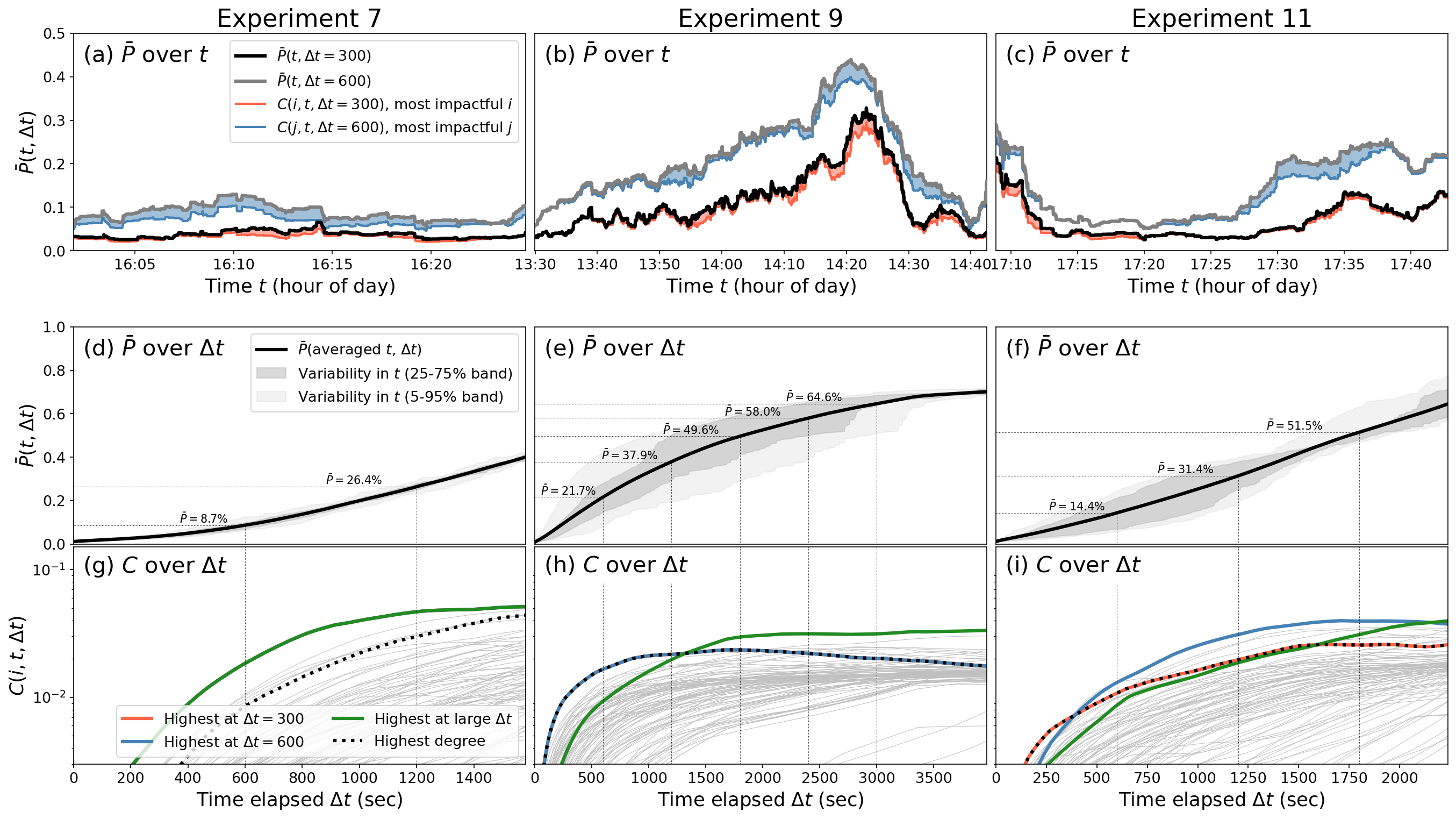}
    \caption{Evolutions of $\bar{P}$ and $C$ of experiments 7 (left column), 9 (middle column) and 11 (right column).
    \textbf{Panels (a)-(c)}: Evolution of $\bar{P}(t, \Delta t)$ over time $t$ at $\Delta t = $300 seconds (black/red) and $\Delta t = $600 seconds (gray/blue). We also show $\bar{P}_{-i}(t, \Delta t)$ (red and blue curve, respectively) and $C(i, t, \Delta t)$ (red and blue area) for the most impactful agents $i$ at these levels of $\Delta t$.
    \textbf{Panels (d)-(f)}: Evolution of $\bar{P}(t, \Delta t)$ over elapsed time $\Delta t$, averaging over $t$, for the same experiments 7, 9 and 11. For a few values of $\Delta t$, we indicate the value of $\bar{P}(t, \Delta t)$.
    \textbf{Panels (g)-(i)}: Evolution of $C(i, t, \Delta t)$ for each agent $i$ (in gray). Specific agents are denoted in colors: the agents with the highest $C$ at $\Delta t = $300 (red) and 600 seconds (blue) as in panels (a)-(c), the agent with the highest $C$ at a near-maximum $\Delta t$ (i.e., 90\% of the maximum time elapsed) in green, and the agent with the highest static contact degree as a black dashed line. Note that in panel (g), the red, blue and green lines overlap (only green shows), and in panel (h), the red and blue overlap (only blue shows).}
    \label{fig:3}
\end{figure}

In Fig.~\ref{fig:3}, in order to illustrate differences among the experiments, we show the time evolution of the system property $\bar P$ and the contact sequence centrality $C$ (for specified agents) in three of these experiments: 7, 9 and 11. The conditions behind these experiments are quite different as shown in Tab.~\ref{tab:exps}: experiment 11 contained no interventions at all, while direction and buzzer feedback interventions are differently applied in experiments 7 and 9. Importantly, we use the experiments to highlight several aspects of $\bar{P}$ and $C$.

First, in the top row, we show the evolution of $\bar{P}(t,\Delta t)$ --- see definition in Eq. (\ref{e2}) --- as a function of (the starting time) $t$. A parameter that needs to be chosen is $\Delta t$, modulating the time interval across which the tracing between agents is done (see Fig.~\ref{fig:1}(b) for an example). This choice requires special attention, because while some agents may have most impact in the short run, others may have more impact on the long run. In other words: the individual differences in impact on contact sequences, the main topic of this paper, are dependent on the `time scale' $\Delta t$ we are interested in (we will come back to this later). In Fig.~\ref{fig:3}(a)-(c), we choose $\Delta t\,=$ 300 and 600 seconds. For example, the black curve represents the average fraction of unique agents in the time interval $(t,t+300$ seconds) in the contact sequence of any agent at time $t$. High values at time $t$ (may) indicate that large and frequent events are taking place in the interval $(t, t+300\text{ sec})$, involving many unique mingling people, as opposed to mingling only within specific bubbles or clusters. Clearly, the grey curve ($\Delta t = 600$) is always above the black curve ($\Delta t=300$), which is expected, as over a longer time span $\bar{P}$ has more time to grow. Visual inspection of the panels (a)-(c) immediately reveals a number of differences in the evolution of $\bar{P}(t,\Delta t)$ among the different experimental settings: while experiment 7 remains rather flat at low values, experiment 9 contains a build-up towards much higher values (up to 0.30 for $\Delta t = 300$ sec), and experiment 11 lies in between the two and seems to be split in alternating phases of highs and lows. Without going into the specifics of these experiments, the shapes of the $\bar{P}$ curves can be interpreted as follows. A constant, flat curve (like in experiment 7) indicates that the interaction configuration is such that the potential of any spreading phenomenon does not vary much in time: if a disease or rumor would start at the start of the time interval, or in the middle, it would spread with roughly equal speed. However, a lot of variation in $\bar{P}$ (like in experiment 9) indicates that the structure of interactions and the temporal graph itself is different when starting at different time points. Sometimes, it promotes tracing many agents through contact sequences (e.g., at 14:20, we have that $\bar{P} = 40\%$ in 10 minutes) by means of very frequent events with randomized attendees, which may imply that the system is vulnerable to spreading dynamics like disease transmissions and rumor spreading. In other cases, the temporal network structure only contains narrow contact sequences, tracing only few agents to each other (e.g., at 13:30, we have that $\bar{P} = 5\%$ in 10 minutes), which is the case in a well-segregated agent population.

In Fig.~\ref{fig:3}(a)-(c), to mark the effect of individual agents in the system, we also plot $C$ after fictitiously removing the `most impactful agent' at both levels of $\Delta t$ plotted here: 300 and 600 seconds, in red and blue, respectively. Note that there are different ways to define the `most impactful agent', and the correct choice is depending on the application. In particular, the ranking of agents in terms of their impact on the contact sequences is in reality dependent on both $t$ and $\Delta t$, i.e., who is `most impactful agent' varies over time. For illustration purposes we choose to average all $C$ values over $t$ and look at the agent with the highest value of $C$ at fixed, chosen values of $\Delta t$ (in other panels we use the same procedure). In other words, our definition of `most impactful agent' is the agent that at a chosen value of $\Delta t$, has the highest value of $C$, averaged over the possible time starting points $t$. In the panels (a)-(c), the reduction in $\bar{P}$ (marked by the red and blue areas) if we would remove these high-impact agents remains relatively invariant over time $t$ in experiment 7, while in experiments 9 and 11, there are clear moments in time where the most impactful agents seem to have their highest impact: for experiment 9, this is near the peak, and for experiment 11, most impact is found either at the start, or around 17:27-17:37. The agents denoted in red and blue are not necessarily the same agent (see below).

While in the upper row, we see how $\bar{P}$ evolves over time, indicating where contact sequences promoted any spreading dynamics most, we now look at how $\bar{P}$ grows with $\Delta t$ in panels (d)-(f), which reflects how many people you can trace via contact sequences within a given time interval size $\Delta t$. These are monotonically increasing curves: taking a larger time window allows the system reach equal or more agents, but never fewer. Cross-experimental differences can be observed in the slope at which $\bar{P}$ is increasing with $\Delta t$: at $\Delta t=10$ minutes, while $\bar P$ only reaches 8.7\% in experiment 7, it reaches 14.4\% in experiment 11, and even 21.7\% in experiment 9. These differences based in the topology of the respective temporal networks, but may have implications on potential dynamics on top. For example, the fact that the contact sequences in experiment 9 reach many more agents in the same time interval as experiment 7, implies that spreading phenomena like disease transmission or rumor spreading may be more fast-paced in the setting of experiment 9. There are several considerations to be made when concluding this, as discussed in 4.1. While the black lines show averages over $t$, the increase of $\bar{P}$ with $\Delta t$ varies in time -- sometimes this growth is faster than in other moments -- which is shown in the grey shaded areas. Note that the variations around the black curve are much larger in experiments 9 and 11 than in experiment 7, also seen in panels (a)-(c).

%% ==================================== %%
\subsection{Relating $C$ to existing network metrics}
%% ==================================== %%

The bottom panels (g)-(i) in Fig.~\ref{fig:3} show $C(i, t, \Delta t)$, which is the effect of the removal of agent $i$ on the black curves shown in panels (d)-(f). The curves of all agents $i$ are shown in gray, but we focus on four aspects: the agent with the highest $C$ for $\Delta t = 300$ in red, the same for $\Delta t = 600$ in blue, the same for $\Delta t = 90$\% of the total time in green, and the agent with the highest degree in the aggregated contact network in the interval $(t,t+\Delta t)$ in black dashed lines. Note that we scanned $C$ based on a few chosen $\Delta t$ values to highlight particular agents (in red, blue and green), but then plotted these agents over the whole spectrum of $\Delta t$ values (on the horizontal axis) -- `their' $\Delta t$ value is merely to label them and does not play any further role. We henceforth refer to these agents as the `red', `blue' and `green' agents. The reason for plotting the highest-degree agent (black dashed) is to assess whether having a high degree in the aggregated contact network is be related to the most impactful agents and the various time scales $\Delta t$. The answer clearly depends on the experiments: in experiment 7, there is a single curve clearly above all others on all time scales $\Delta t$, which means that the red, blue and green agents overlap. With a clear separation to the second and third places, the highest degree agent is not even close to having most impact on larger time scales (note that the vertical scale is logarithmic, which creates a natural convergence of high-impact curves to the right), albeit that this agent remains is in the top 10\% of agents with highest $C$ values. In experiment 9, the place for the highest $C$ switches between agents. Up to values of $\Delta t$ of about 1300 seconds, a particular agent dominates $C$ (overlapping blue and red), which also happens to be the agent with the highest degree. However, at $\Delta t = 1300$ sec, the green agents takes over the dominance in $C$, who had a modest impact before that. Such separation reveals the dependency on time scale when assessing the impact of a specific agent on contact sequences. In particular, it confirms the hypothesis that aggregated properties do contain most information for contact sequence centrality at small values of $\Delta t$, while they lose their relevance at larger values of $\Delta t$. This is highly pronounced in experiment 11. At short time scales, the red curve marks the highest-$\Delta \bar{P}_i$ individual, which also happens to be the highest-degree agent. Around $\Delta t \approx 400$ sec, this individual is surpassed by another (marked in blue), which dominates the spectrum of $\Delta \bar{P}_i$ at most time scales, but is caught up by yet another (in green) at the largest time scale, where the red/highest-degree agent is clearly having a lower impact than many other agents.

\begin{figure}[!h]
    \centering
    \includegraphics[width=1\linewidth]{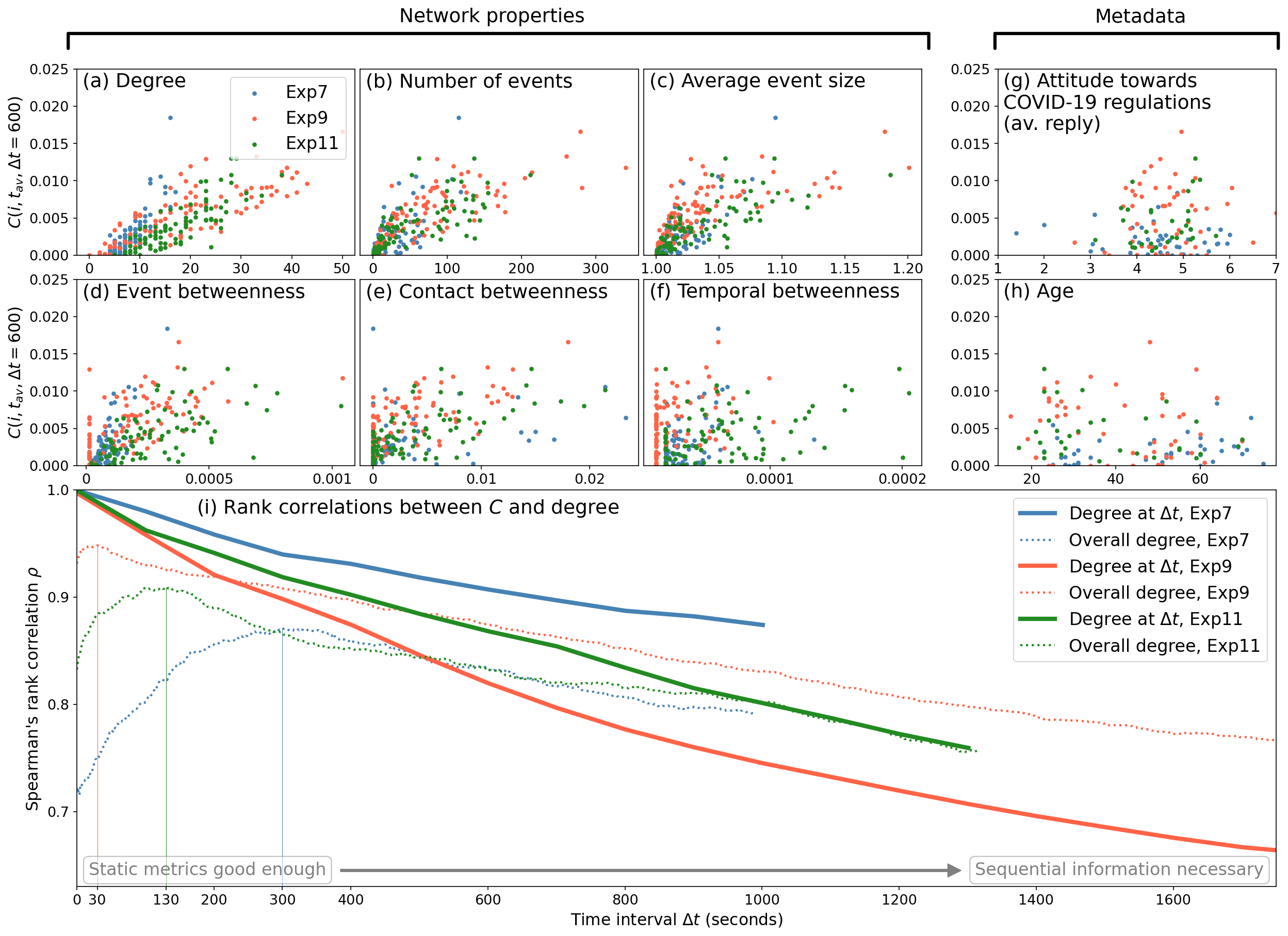}
    \caption{\textbf{Panels (a)-(f)}: Relation between agent's impact on $C$ and various network properties. Different colors denote different experiments.
    \textbf{Panels (g)-(h)}: Same as in panels (a)-(f), but for two metadata metrics: the inferred attitude of the agent with respect to COVID-19 measures based on questionnaire data, and the agent's age.
    \textbf{Panel (i)}: Spearman's rank correlation coefficients between contact degree and $Ci$ for different values of $\Delta t$ for each of the three experiments. Thick lines indicate the correlation coefficients when using the average degrees of the agents when looking at interval sizes $\Delta t$ (as marked by the horizontal axis), while the dotted lines use the overall degree, independent of $\Delta t$. Per experiment, $\Delta t$ values of up to half of the experiment's total time interval are used, to keep enough initialization times $t$ to average over.}
    \label{fig:4}
\end{figure}

Further, Fig.~\ref{fig:4}(a)-(f) show the relationships between $C(i, t, \Delta t)$ at a fixed value of $\Delta t$ (averaged over $t$) --- denoted by $C(i, t_{av}, \Delta t)$ --- and six existing network metrics. On the one hand, in panels (a)-(c) we show three such scatterplots: $C(i, t_{av}, \Delta t)$ versus the degree of the agent $i$ in the aggregated network, obtained from the aggregated contact network; the number of (non-individual) events the agent participates in; and the average size of the events (including those including only a single disconnected agent) in the time interval $(t,t+\Delta t)$. Being obtained from the aggregated network, these are all `static' network properties: they do not provide information on the sequences of interactions. For example, the degree is obtained from the static contact network and frequency or sequences are not taken into account. Likewise, the number of non-individual events an agent has been participating in does not necessarily reflect a wide range of agents --- all participated events may be containing only the same people, and the average event size reflects how often the people participates in individual events (i.e., isolated from the rest of the agents). All three metrics (especially the [fully aggregated] degree) correlate relatively well with $C$ for smaller values of $\Delta t$, across all three experiments (in colors). Still, there is quite some spread: agents with the highest $C$ are not the ones with the highest network metrics. Also, while experiment 7 has generally smaller degree, the scatter in the $C$ values seems to be equal to the other two experiments. On the other hand, panels (d)-(f) describe three forms of betweenness: event, contact and temporal betweenness (see Sec. 2.2). A clean relation between $C$ and temporal or event betweenness is barely visible.

Also relevant to the art fair experiment are metadata obtained from questionnaires (for details, see Ref. \cite{Tanis2020}). We aggregated questions related to the attitudes of the individuals towards the interventions and dangers of the COVID-19 pandemic that was present during the time of the experiment (August 2020), involving questions like "Do you adhere to the 1.5 meter distance rule as issued by the Dutch government?". Participants' reply on a scale from 1 to 7, where a lower score indicates more skepticism on the pandemic and less strict following of the Government imposed COVID-19 rules, while a higher score means the opposite. The average over 22 of such questions is shown at the horizontal axis in panel \ref{fig:4}(g). Clearly, there are no significant differences between the experiments, and no relation is visible with $C$. In panel \ref{fig:4}(h), we also plot the age of these participants, also yielding no significant relationship with the impact on spreading capability. Not all agents filled in the questionnaires however, resulting in less dots in the metadata panels than in panels (a)-(f).

Putting together Figs.~\ref{fig:3}(g)-(i) and Fig.~\ref{fig:4}(a)-(f), several relations between network metrics and $C$ can be inferred, but any such relation would be subject to the time scale $\Delta t$: for small values of $\Delta t$, the static degree may be well suited to describe an agent's impact: having more unique direct contacts means that the agent, potentially, have been in contact with many other agents, which directly affects $\bar{P}$. In turn, therefore, removing that agent would result in a higher $C$. At higher time scales, however, the effect of sequences becomes more important. Agents with (potentially) smaller initial impact and contact degree may have a larger impact on larger time scales because they serve as conduits of contact for large bubbles or other more subtle temporal structures. This is visible in Fig.~\ref{fig:3}(g)-(i). To explicitly test the effect of time scale on whether traditional network metrics are sufficient, we show the the Spearman's rank correlation coefficient between the contact degree and $C$ as a function of $\Delta t$ in Fig.~\ref{fig:4}(i). The overall degree (independent from $\Delta t$) is also correlated to $C$, shown in dotted lines. A clear decay of this correlation with $\Delta t$ is visible, reflecting that for small $\Delta t$, static metrics like the contact degree are sufficient, but for larger time scales, the sequential information becomes important, which motivates a more complex metric like contact sequence centrality $C$.

%% ============================================================================ %%
%% Main text - Section 4
%% ============================================================================ %%
%% ============================================================================ %%
\section{Discussion and Conclusion}
\label{sec:4}
%% ============================================================================ %%

In this paper we have developed and demonstrated a methodology to quantify agent impact on contact sequences in social interactions from the perspective of spreading phenomena. Spreading phenomena are versatile in their nature and types; however, one common element they share is that they all require individual agents as carriers of the entity, and an agent in possession of the concerned entity can pass it on to others that do not have it, leading to the natural expectation that the temporal topology of agent interactions --- who interacts with whom, when and in which sequence --- will have a profound influence on the spreading dynamics. 

In order to quantify this effect, we have proposed a new measure for individual differences in their impact on contact sequences. The measure is obtained by porting all contact information --- while preserving the sequences of contacts --- into an event graph, and tagging the connections between events by agents' temporal paths. From the event graph, we determine the metric $\bar{P}(t, \Delta t)$, which is the average fraction of agents at time $(t+\Delta t)$ traceable from any agent at time $t$. Likewise, contact sequence centrality $C(i, t, \Delta t)$ represents an agent $i$'s impact on the contact sequence in terms of how strongly $\bar{P}$ changes when all temporal links belonging to agent $i$ are removed.

For event mapping, we have used the convention that at every (integer) snapshot the agents' contact network within an event is a complete one. For example, for the agents' contact network snapshot at time $t$, agents 1-and-2 and 2-and-3 may have been recorded to directly interact, but not 1-and-3: in such a case event mapping introduces a direct interaction between agents 1-and-3 that are not originally present in the empirical data. While this is a subtle issue discussed elsewhere in detail \cite{Dekker2021temp}, we note the value of $C$ is dependent on the choices made in this convention. This is not necessarily a drawback for the concept of contact sequence centrality that we have developed here, but may limits its applicability depending on the time-scale of transmission of a quantity of interest. For example, imagine that the time-scale for transmission is very short relative to the typical duration of an event. Then, in the above example event involving three agents 1, 2 and 3, it will not matter whether 1 and 3 interact directly or not: agent 1 will effectively be in direct contact with agent 3.

Note also that we have calculated both $\bar{P}$ and $\bar{P}_{-i}$ forward in time. This is an obvious choice for spreading dynamics, but for other applications, calculating these quantities backward in time would be more appropriate. For example, when interested in the vulnerability of a specific vendor in a supply chain, the more time-backward dependence the vendor has, the more it will be vulnerable to random failure. In other words, the choice of the direction of time for calculating $\bar{P}$ and $\bar{P}_{-i}$ depends on the application. Note also that the removal of a specific agent in our calculations is merely a conceptual construct to assess its impact on the contact sequence --- the actual removal of this agent may have unforeseen effects on the topology of the temporal network that cannot be captured in a conceptual framework.

Contact sequence centrality is a purely topological property, reflecting properties of connections. In other words, agents with high levels of $C$ can be treated as what are colloquially referred to as (behavioral) \textit{super-spreaders} \cite{Zhang2019, Kitsak2010, Ahajjam2018}, being more central to the contact sequences. Our work serves to highlight the fact that super-spreaders are not necessarily those individuals that have most contacts: as shown in Fig.~\ref{fig:3}(g)-(i), we see that the highest-degree agent starts losing its dominance in $C$ when looking at larger time scales. In particular, increasing the time interval $\Delta t$ makes the degree less and less determinant for the $C$ values, as shown in Fig.~\ref{fig:4}(i). What sets our metric apart from aggregated (and static) and various temporal metrics is that the exact sequence of the interactions matter: in the example of Fig.~\ref{fig:1}(c), removing agent 3 removes an edge early on in the time series, which has cumulative effects later on. Moreover, reversing the direction of time would yield the same values of all static metrics like degree, and also for temporal betweenness for the aggregated networks, even though it would fundamentally change the values $C$. Note that our approach had a specific focus on identifying an agent's spreading potential, while many other related approaches exist, each with own scope and (dis)advantages \cite{hafiene2020influential}. For example, Yu et al. (2020) combines network embedding with machine learning to assess a node's importance in terms of spreading as evaluated by a SIR-model, which contrasts with our approach relying only on topology \cite{Yu2020}.

Finally, it is important to realize that topology of the contact sequences alone is not enough for the behavior-epidemiology interface, or for that matter for any dynamical process playing out on top of the temporal network; the process itself may have its own inherent time-scale parameters. Let us elaborate a bit on this point using an example like pathogen spreading, which comes with its own time-scales, such as incubation period, or the time required for an infected agent can itself be infectious. On a given school day, all students in a school may be sequentially traceable to each other as measured by our metrics: $P_i=1$ for all students $i$, and $\Delta t = 1$ day. However, the disease may only spread to students directly linked to a patient zero $i_0$, simply because secondary contacts do not matter because of the incubation time of COVID-19: the direct contacts were not infectious yet. Given this, it is natural to expect that the eventual dynamics of pathogen spreading will be a combination all time-scales involved, including event frequency in the temporal network.

Indeed, to make the current approach relevant for epidemiology, the observation period of contacts among the agents needs to be sufficiently longer than the incubation period, and the time-lag between being infected and being infectious, bringing us to an issue of practicality. While the event mapping is a relatively quick process to compute, depending on the amount of time steps $T$, the calculation of $\bar{P}$ scales with an extra factor $N$ because the contact sequence tracking [e.g., as in Fig.~\ref{fig:1}(b)] has to be done for every agent. However, when computing the $C$ for every agent $i$, yet another factor $N$ is brought in, as all calculations have to be done fully over when removing $i$. Hence, the computational complexity of $C$ is of the order $\mathcal{O}(TN^2)$. This means, given that the contact networks of the agents are sampled at 1-second interval, that for making the current method applicable for COVID-19 epidemiology (with an incubation time of about 5.2 days \cite{Li2020inc} and infectiousness starting time of 12.3 days \cite{He2020}), further course-graining methods to preprocess the network in the the temporal domain will need to be developed.

In the current paper, we have analyzed an application inspired by the COVID-19 pandemic, as the structure and dynamics of contact networks have proved pivotal to forge a conceptual link between human behavior and virus spread. However, our understanding of human behavior is still limited by a lack of adequate measurement techniques and modeling frameworks. The work presented in this paper showcases an extension of approaches to data analysis, in which experimental manipulations can be tracked to a level of detail that was not previously available. As is the case for any data analytic method, the approach requires existing high-resolution data. Currently, this limits the possibility of applying these calculations to country-wide scales, typically associated with spreading processes such as epidemics. The contribution of this work primarily lies in the characterization of local environments such as supermarkets and museums --- it reveals system properties rather than laying out the full spreading infrastructure.

On the basis of the growing number of experimental studies on the topic, this system-characterizing methodology may become available in many situations, which may allow for policy making through the anticipation of spreading vulnerability in future situations. Potentially, the methodology is also used in simulation software that one could use to run through different possible floor plans of an event such as the art fair described in this paper. Policymakers would be in a position to assess different scenarios while planning events. While such techniques are not yet available, we hope that detailed insights into human behavior as obtained through the current analysis may contribute to their development.

In our view, further development of technology and methodology to obtain and analyze direct measures of behavioral contact networks is essential to advance our understanding of human behavior, and to improve the resilience of society in dealing with pandemics. 

\FloatBarrier

%% ============================================================================ %%
%% References
%% ============================================================================ %%

\bibliographystyle{bibstyle}
%% Added to format the references better
\let\oldthebibliography\thebibliography
\let\endoldthebibliography\endthebibliography
\renewenvironment{thebibliography}[1]{
  \begin{oldthebibliography}{#1}
    \setlength{\itemsep}{0em}
    \setlength{\parskip}{0em}
}
{
  \end{oldthebibliography}
}

%\bibliography{literature_main}

%\bibliography{literature_main}

%% ============================================================================ %%
%% Acknowledgements
%% ============================================================================ %%

\section*{Acknowledgments}
Figure~2 is reprinted from \cite{Tanis2020}, for which we acknowledge the authors and thank Matthijs Immink (https://matthijsimmink.com) for the photo of the art fair.

\textbf{Funding:} This work is part of the project `Evidence-based effective monitoring and control of COVID-19 after the initial outbreak', which has been funded by the Netherlands Organization for Health Research and Development (ZonMw) with project number 10430022010001. This work is also part of the research programme `Improving the resilience of railway systems' with project number 439.16.111. MMD was supported by this research project, which is financed by the Dutch Research Council (NWO) and co-financed by Nederlanse Spoorwegen (NS) and ProRail.

\textbf{Author contributions:} All authors contributed to the research conceptualization. MMD and DP conceived the mathematical principles, with help of JO. MMD performed the data analysis and application to the art fair, with help of TB, FD and DB. MMD wrote first draft of the manuscript. All authors reviewed the final text.

\textbf{Competing interests:} The authors declare no competing interests.

\textbf{Data and materials availability:} The code for calculating the contact sequence centrality will be made available upon publication. Details about the data of the art fair experiments can be found in Ref. \cite{Tanis2020}.

\section*{Supplementary Materials}
SI A - Supplementary Plots

%% ============================================================================ %%
%% Figure legends
%% ============================================================================ %%
\newpage
\setcounter{figure}{0}
\captionsetup[figure]{font=normalsize,labelfont={normalsize, bf}}

\section*{Figure legends}

\begin{figure*}[!h]
    \centering
    \caption{Illustration of the event mapping and an individual's contact sequence centrality using a toy example. All throughout this figure, individual agents are in squares (and numbers), events are in circles. \textbf{Panel (a)}: In a temporal network, the time evolution of the contact graph (middle) captured in a sequence of snapshots. Contact data over the four snapshots can be aggregated to obtain a static (and aggregated) contact network (left), but in doing so, the sequential information is lost. We term the connected network components within each snapshot `events', they are seen as nodes in an event graph (right). In an event graph the event-to-event trajectories of agents are the links. Colors of circles and snapshots indicate time from blue to red. \textbf{Panel (b)}: Example of a contact sequence when tracing all events and subsequent agents back to agent 5 at $t=0$. Starting time $t$, time elapsed $\Delta t$, the percentage unique agents [$P(5)$], traceable to 5 and the average percentage of agents in the contact sequence $\bar{P}$ are indicated.
    \textbf{Panel (c)}: Same as (b), but after removal of agent 3, resulting in a `perturbed' contact sequence of agent 5. The (now absent) links that were associated with agent 3 are in dashed green arrows, and the events that are consequentially not traceable to 5 anymore are colored green. The associated (lowered value of) $P_{-3}(5)$ is shown. Averaging over all agents provides $\bar{P}_{-3}$, and allows us to calculate the \textit{Contact Sequence Centrality} of agent 3: $C(3) = \bar{P} - \bar{P}_{-3}$, which is clearly marked in the bottom panel. For abbreviation, we excluded the $t$ and $\Delta t$ between brackets when showing values of $P$, $\bar{P}$ and $C$.}
    %\label{fig:1}
\end{figure*}

\begin{figure*}[!h]
    \centering
    \caption{The art fair on Amsterdam on the first day with bidirectional walking directions (top) and schematic layout of the location (bottom). The art fair consisted of 28 stands, spanning 1,080m$^2$. Visitors entered on the left of the plan in the bottom panel, where also the access point was placed. Figure adapted and reprinted with permission from Ref. \cite{Tanis2020}.}
    %\label{fig:2}
\end{figure*}

\begin{figure*}[!h]
    \centering
    \caption{Evolutions of $\bar{P}$ and $C$ of experiments 7 (left column), 9 (middle column) and 11 (right column).
    \textbf{Panels (a)-(c)}: Evolution of $\bar{P}(t, \Delta t)$ over time $t$ at $\Delta t = $300 seconds (black/red) and $\Delta t = $600 seconds (gray/blue). We also show $\bar{P}_{-i}(t, \Delta t)$ (red and blue curve, respectively) and $C(i, t, \Delta t)$ (red and blue area) for the most impactful agents $i$ at these levels of $\Delta t$.
    \textbf{Panels (d)-(f)}: Evolution of $\bar{P}(t, \Delta t)$ over elapsed time $\Delta t$, averaging over $t$, for the same experiments 7, 9 and 11. For a few values of $\Delta t$, we indicate the value of $\bar{P}(t, \Delta t)$.
    \textbf{Panels (g)-(i)}: Evolution of $C(i, t, \Delta t)$ for each agent $i$ (in gray). Specific agents are denoted in colors: the agents with the highest $C$ at $\Delta t = $300 (red) and 600 seconds (blue) as in panels (a)-(c), the agent with the highest $C$ at a near-maximum $\Delta t$ (i.e., 90\% of the maximum time elapsed) in green, and the agent with the highest static contact degree as a black dashed line. Note that in panel (g), the red, blue and green lines overlap (only green shows), and in panel (h), the red and blue overlap (only blue shows).}
    %\label{fig:3}
\end{figure*}

\begin{figure*}[!h]
    \centering
    \caption{\textbf{Panels (a)-(f)}: Relation between agent's impact on $C$ and various network properties. Different colors denote different experiments.
    \textbf{Panels (g)-(h)}: Same as in panels (a)-(f), but for two metadata metrics: the inferred attitude of the agent with respect to COVID-19 measures based on questionnaire data, and the agent's age.
    \textbf{Panel (i)}: Spearman's rank correlation coefficients between contact degree and $Ci$ for different values of $\Delta t$ for each of the three experiments. Thick lines indicate the correlation coefficients when using the average degrees of the agents when looking at interval sizes $\Delta t$ (as marked by the horizontal axis), while the dotted lines use the overall degree, independent of $\Delta t$. Per experiment, $\Delta t$ values of up to half of the experiment's total time interval are used, to keep enough initialization times $t$ to average over.}
    %\label{fig:4}
\end{figure*}
\end{document}